\documentclass[onecolumn,showpacs,preprintnumbers,amsmath,amssymb]{revtex4}

\usepackage{epsf}
\usepackage{graphicx}  % Include figure files
\usepackage{dcolumn}   % Align table columns on decimal point
\usepackage{bm}        % bold math

\newcommand{\be}{\begin{equation}}
\newcommand{\en}{\end{equation}}
\newcommand{\bea}{\begin{eqnarray}}
\newcommand{\ena}{\end{eqnarray}}

\newcommand{\g}{\gamma}

\begin{document}

%\preprint{KASI/0000-2008}

\title{Probing the nature of cosmic acceleration}

 \author{Hongsheng Zhang\footnote{Electronic address: hongsheng@kasi.re.kr} }
 \affiliation{ Korea Astronomy and Space Science Institute,
  Daejeon 305-348, Korea }
 \affiliation{ Department of Astronomy, Beijing Normal University,
Beijing 100875, China}

\author{Heng Yu\footnote{Electronic address:
 gerry@mail.bnu.edu.cn}}
  \affiliation{ Department of Astronomy, Beijing~~ Normal University,
  Beijing 100875, China}

\author{Hyerim Noh \footnote{Electronic address: hr@kasi.re.kr} }
 \affiliation{
 Korea Astronomy and Space Science Institute,  Daejeon 305-348, Korea }
 \author{Zong-Hong Zhu\footnote{Electronic address:
zhuzh@bnu.edu.cn}}
  \affiliation{ Department of Astronomy, Beijing Normal University,  Beijing 100875, China}

\begin{abstract}
 The cosmic acceleration is one of the most significant cosmological discoveries
 over the last century. The two categories of explanation are exotic component (dark energy) and
 modified gravity. We
 constrain the two types of model by a joint analysis with
 perturbation growth and direct $H(z)$ data. Though the minimal $\chi^2$ of the $\Lambda$CDM is almost the same as that of DGP, in the sense of consistency
we find that the dark energy ($\Lambda$CDM) model is more favored
through a detailed comparison with the corresponding parameters
fitted by expansion data.

\end{abstract}


\maketitle

\section{Introduction}
 The acceleration of the universe is one of the most significant cosmological discoveries
 over the last century \cite{acce}. Various explanations of this acceleration have been
 proposed, see \cite{review} for
 recent reviews with fairly complete lists of references of different models.
 However, although fundamental for our understanding of the universe, its nature remains as a completely
 open question nowadays.

 There are two main categories of proposals. One is that the
 acceleration is driven by some exotic matter with negative pressure, called dark energy.
 The other suggests that general relativity fails in the present
 Hubble scale. The extra geometric effect is responsible for the acceleration.
  Surely, there are some proposals which mix the two
 categories. Mathematically, in the dark energy model we present
 corrections to the right hand side of Einstein equation (matter part), while the
 correction terms appear in the left hand side of Einstein equation (geometric part).

 $\Lambda$CDM model is the most popular and far simple dark energy
 model, in which vacuum energy with the equation of state (EOS)
 $w=-1$ accelerates the universe. From theoretical considerations
 and by observational implications, people put forward several other
 candidates for dark energy, such as quintessence ($-1<w<-1/3$), phantom
 ($w<-1$), etc. Also there are many possible corrections to the
 geometric part of the theory. One of the leading modified gravity
 model is Dvali-Gabadadze-Porrati (DGP) model \cite{dgpmodel}, for a
 review, see \cite{dgpre}. In the DGP model, the bulk is a flat Minkowski spacetime, but an induced
 gravity term appears on a tensionless brane.
 In this model, gravity appears
 4-dimensional at short distances. But, at a distance larger
 compared to some freely adjustable crossover scale $r_0$ it is altered  through the
 slow evaporation of the graviton off our 4-dimensional brane world
 universe into an unseen, yet large, fifth dimension.

 We should find the correct, at least exclude the incorrect models in the model
 sea. The first step is to discriminate between dark energy and
 modified gravity, whose nature are completely different. To
 construct a model simulating the accelerated expansion is not very
 difficult. That is the reason why we have so many different models.
  Recently, some suggestions are presented that growth
 function $\delta(z)\equiv\delta\rho_m/\rho_m$ of the linear matter
 density contrast as a function of redshift $z$ can be a probe to
  discriminate between dark energy and
  modified gravity \cite{dgpgrow, growth} models. The growth function can break
  the degenerations between dark energy and modified gravity models which share the same
  expansion history.

  There is an  approximate relation between the growth function
  and the partition of dust matter in standard general relativity \cite{peeb},
   \be
   \label{f}
 f\equiv\frac{d\ln\delta}{d\ln a}=\Omega_m^\gamma,
   \en
 where $\Omega_m$ is the density partition of dust matter, $a$
 denotes the scale factor, and $\g$ is the growth index. This relation is a
 perfect approximation at high redshift region. Also, it  can be
 used
 in low redshift region, see for example \cite{stein}. It is shown that the relation
 ($\ref{f}$) is also valid in the case of modified gravity theory \cite{dgpgrow}. The
 theoretical value of $\g$ for $\Lambda$CDM model is $6/11$ \cite{growth, stein},
 while,
 for spatially flat DGP model is $11/16$ \cite{dgpgrow}. The observation data of
 perturbation{ growth} are listed in Table 1.
 {\large
 \begin{table}
 \begin{center}
 \begin{tabular}{ccccc} \hline\hline
 $z$ &  ~~~~~$f_{obs}$ & Reference\\ \hline
 $0.15$ &~~~~~ $0.51\pm 0.11$ & \cite{hawk}\\
 $0.35$
 & ~~~~~$0.70\pm 0.18$ & \cite{teg}\\
 $0.55$ &~~~~~ $0.75\pm 0.18$ & \cite{ross}\\
 $0.77$ &~~~~~ $0.91\pm 0.36$ & \cite{guzz}\\
 $1.4$
 &~~~~~ $0.90\pm 0.24$ & \cite{angl}\\
 $3.0$ &~~~~~ $1.46\pm 0.29$ & \cite{dona}\\
 \hline\hline
 \end{tabular}
 \end{center}
 \caption{\label{tab1} Observed perturbation growth as a function of redshift $z$, see also \cite{weihao} }
 \end{table}}

 It is shown that $\Lambda$CDM model is consistent with the current
 growth data \cite{favour}. The data seem to weaken a spatially flat DGP model, whose $\g=11/16$.
 However, it is found that the growth index $\g$ is $4/7$ in a non spatially flat DGP model \cite{weihao}, which is
 very closed to the index of $\Lambda$CDM. Thus, the DGP model may be still consistent with current growth data.

 In this article we take a different strategy. The previous works
 were concentrated on the limit of the growth index  and made some approximations
 on it
 (often the high $z$ limit was assumed and an approximation was made at linear order),
   in which only approximate asymptotic
  value of $\g$ can be obtained. In fact, the perturbation growth
  $f$ is a variable with respect to $z$, as displayed in Table 1. By using these growth data we
  constrain the parameters in $\Lambda$CDM model and DGP model,
  respectively.

 The other one which is
 very useful but not widely used in model constraint data is the set of direct $H(z)$.
 $H(z)$ is derived by a newly
 developed scheme to obtain the Hubble parameter directly
 at different redshift \cite{h(z)}, which is based on a method
 to estimate the differential ages of the oldest galaxies
 \cite{age}. By using the previously released data \cite{aid},
 Simon {\it et al.} obtained a sample of direct $H(z)$ data in the interval $z\in
 (0,1.8)$~\cite{simon}, just as the same interval of the data of
 luminosity distances from supernovae. For the present sample of growth
 data derived with the assumption of the expansion behaviors of the universe, we will present
 joint fittings to obtain the constraints on the $\Lambda$CDM and
 DGP, respectively. Then, through comparing with allowed regions by expansion constraint using different
  observations, including supernovae (SN), cosmic microwave background (CMB), baryon acoustic oscillations (BAO) etc., we examine which
  model is more self-consistent.

  This article is organized as follow:  In the next section we
  construct
  the evolution equation for $f$ in a very general frame. In
  section III, by using the growth data and $H(z)$ data we present the parameter constraints of $\Lambda$CDM
  and DGP, respectively. Our conclusion and some discussions appear
  in the last section.

 \section{the evolution equation for the growth function $f$}
 We consider a mixed model in which dark energy drives the universe
 to accelerate
 in frame of modified gravity. For  FRW universe in modified
 gravity, the Friedmann equation can be written as,
 \be
 H^2+\frac{k}{a^2}+h(a,\dot{a},\ddot{a})=\frac{8\pi
 G}{3}(\rho_m+\rho_e),
 \en
 where $H$ denotes the Hubble parameter, $h$ comes from the
 corrections to general relativity. $\rho_m$ and  $\rho_e$ represent the density
 of dust matter and the exotic matter, respectively. A dot implies the derivative with respect to cosmic time $t$.
  Comparing with the corresponding Friedmann equation in
 standard general relativity, we obtain the density of effective
 dark energy,
  \be
  \rho_{de}=\rho_e-\frac{3}{8\pi
 G}h.
  \en
  Here we call $h$ geometric sector of dark energy. The behavior of the
  effective dark energy has been separately discussed in some
  previous works. For example, it is investigated in detail in a
  modified gravity model where a four
 dimensional curvature scalar on the brane and a five dimensional
 Gauss-Bonnet term in the bulk are present \cite{self}.

 For any modified gravity theory, Bianchi identity is a fundamental
 requirement. Using the continuity equation of the dust matter and the Bianchi identity, we derive,
  \be
  \dot{\rho_{de}}+3H\rho_{de}(1+w_{de})=0,
  \en
 which yields,
 \be
 w_{de}=-1-\frac{1}{3}\frac{d\ln \rho_{de}}{d\ln a},
 \en
  where $w_{de}$ is the EOS (equation of state) of the effective dark energy.

  After the matter decoupling from radiation, for a region well inside a Hubble radius, the perturbation growth
  satisfies the following equation in standard general relativity \cite{liddle},
  \be
  \ddot{\delta}+2H\dot{\delta}-4\pi G\rho_m \delta=0.
  \label{pertu}
  \en
  It is found that the perturbation equation is still valid in a modified gravity theory if we
  replace the Newton constant $G$ with an effective gravitational
  parameter $G_{eff}$, which is defined by Cavendish-type
  experiment \cite{dgpgrow, modi} (This point may need more studies.). $G_{eff}$ may be time-dependent, for example in
  Bran-Dicke theory and in generalized DGP theory \cite{eff}.

  With the partition functions,
  \bea
  \Omega_m &=& \frac{8\pi G\rho_m}{3H^2},\\
   \Omega_{de}&=&\frac{8\pi G\rho_{de}}{3H^2},\\
   \Omega_k &=& -\frac{k}{a^2H^2},
   \ena
 the perturbation equation (\ref{pertu}) becomes,
  \be
   \left(\ln\delta\right)''
 +\left(\ln\delta\right)'^{2}
 +\left(2+\frac{H'}{H}\right)\left(\ln\delta\right)'
 =\frac{3}{2}\alpha \Omega_m,
  \label{f1}
  \en
  where a prime denotes the derivative with respect to $\ln a$, $\alpha$ is
  the strength of the gravitational field scaled by that of standard
  general relativity,
  \be
  \alpha=\frac{G_{eff}}{G}.
  \en
  $\Omega_m$ and $\Omega_k$ redshift as
  \bea
  \Omega_m  =\frac{ \Omega_{m0}(1+z)^3}{\Omega_{m0}(1+z)^3+\Omega_{k0}(1+z)^2+\frac{8\pi G\rho_{de}}{3H_0^2}}, \\
  \Omega_k = \frac{\Omega_{k0}(1+z)^2}{\Omega_{m0}(1+z)^3+\Omega_{k0}(1+z)^2+\frac{8\pi G\rho_{de}}{3H_0^2}},
  \ena
  where 0 denotes the present value of a quantity.
  The growth function defined in (\ref{f}) is just $({\ln \delta})'$.
  Thus (\ref{f1}) generates,
  \be
  f'+f^2+\left[\frac{1}{2}(1+\Omega_k)+\frac{3}{2}w_{de}(\Omega_m+\Omega_k-1)\right]f=\frac{3}{2}\alpha\Omega_m,
  \label{diff}
  \en
  where we have used

    \be
  \frac{H'}{H}=-\Omega_k-\frac{3}{2}\left[\Omega_m+(1+w_{de})\Omega_{de}\right],
  \en
  and
 \be
  \Omega_m+\Omega_k+\Omega_{de}=1.
 \en
  In $\Lambda$CDM model, we have $w_{de}=-1$ and $\alpha=1$. For the
  self-accelerating branch of DGP model \cite{weihao},
 \be
 w_{de}=\frac{-1+\Omega_k}{1+\Omega_m-\Omega_k},
 \en
 and
 \be
 \alpha=\frac{4\Omega_m^2-4\left(1-\Omega_k\right)^2
 +2\sqrt{1-\Omega_k}\left(3-4\Omega_k+2\Omega_m\Omega_k
 +\Omega_k^2\right)} {3\Omega_m^2-3\left(1-\Omega_k\right)^2+2\sqrt{1-\Omega_k}\left(3-4\Omega_k+2\Omega_m\Omega_k
 +\Omega_k^2\right)}.
  \en
 $r_c$ is another important parameter in DGP model , which is
 defined by the relative strength of five dimensional gravity to
 four dimensional gravity $r_c=G_5/G$. Here $G_5$ is the five
 dimensional gravity constant. We define the partition of $r_c$ as
 \be
 \Omega_{r_c}\equiv 1/(H_0^2 r_c^2).
 \en
 One can derive the following relation from Friemann equation of DGP
 model,
 \be
 1=\left[\sqrt{\Omega_{m0}+\Omega_{r_c}}
 +\sqrt{\Omega_{r_c}}\right]^2+\Omega_{k0}.
 \label{rcmk}
 \en

 With (\ref{diff}) and the observed data of $f$ in Table 1, we can
 fit parameters of the models, either dark energy or modified gravity.

 \section{joint analysis with the growth data and the direct $H(z)$ data}
 In this section we fit $\Omega_{m0}$, $\Omega_{k0}$ in $\Lambda$CDM
 model and $\Omega_{m0}$, $\Omega_{r_c}$ in DGP model with the growth data and
 direct $H(z)$ data by $\chi^2$ statistics,
 respectively. Before fitting with the two sets of data, we present some
 discussions about them.

 The present growth data in Table 1 are far from being precise. We
  have a sample consisting of only six points,
  and the error bars of the growth data are at the same order of
  the growth data themselves. The reason roots in the method by which we derive the data set.

  In the present stage we do not find any absolute probes to the
  perturbation amplitude.  People extract the information of perturbation growth from
  galaxy clustering data through redshift distortion parameter $\beta$ observed in the
  anisotropic pattern of galactic redshifts.  We need the galaxy bias factor $b$
  to get the perturbation growth $f=b\beta$.   The current available galaxy bias can be obtained
  mainly in two ways. The most popular method is to refer to the simulation results of galaxy formations\cite{hawk,teg,guzz,dona}.
  At the present stage the simulations we obtained only in frame of $\Lambda$CDM model.
   The second method to get the galaxy bias depends on the CMB
    normalization \cite{ross}. Also, $\Lambda$CDM model is involved.
      Further, to convert from redshift $z$ to comoving
    distance one should assume a clear relation between distance and redshift.
    For instance Tegmark et al. \cite{teg} adopt a flat $\Lambda$CDM model in which $\Omega_{m0}= 0.25$.
    They also tested that if a different cosmological model is assumed for the
    conversion from redshift to comoving distance, the
    measured dimensionless power spectrum  is varied very slightly
    ($<$1\%) \cite{referee}.

    Hence, we see that people obtain data in Table 1 always by assuming a $\Lambda$CDM
   model. Its reliability may decrease when we use it in the scenarios of other models.
   Fortunately, it is pointed out that this problem can be evaded at least in
   the DGP model since the expansion history in DGP with proper parameters is very similar to
   that of $\Lambda$CDM \cite{weihao}. Since the the growth data are
   derived with some assumptions of expansion history, we should fit
   the model by growth data together with expansion data.

   The direct $H(z)$ data are independent of the data of luminosity
  distances and reveal some fine structures of $H(z)$. They have not
  been widely used in the constraints on dark energy models up to
  now. Here we present a joint fitting of $\Lambda$CDM and DGP with perturbation growth data
  and direct $H(z)$ data.

  We show the sample of $H(z)$ data  in table
 II.

  %==================== table 2 ====================
\begin{table}[]
\begin{center}
\begin{tabular}{c|lllllllll}\hline
 $z$ &\ 0.09 & 0.17 & 0.27 & 0.40 & 0.88 & 1.30 & 1.43
 & 1.53 & 1.75\\ \hline
 $H(z)\ ({\rm km~s^{-1}\,Mpc^{-1})}$ &\ 69 & 83 & 70
 & 87 & 117 & 168 & 177 & 140 & 202\\ \hline
 68.3\% confidence interval &\ $\pm 12$ & $\pm 8.3$ & $\pm 14$
 & $\pm 17.4$ & $\pm 23.4$ & $\pm 13.4$ & $\pm 14.2$
 & $\pm 14$ &  $\pm 40.4$\\ \hline
\end{tabular}
\end{center}
\caption{\label{tabhz} The direct observation data of $H(z)$
 ~\cite{simon}
 .}
\end{table}

Table II displays an unexpected feature of $H(z)$: It decreases
   with respect to the redshift $z$ at $z\sim 0.3$
   and $z\sim 1.5$, which is difficult to be found in the data of supernovae since the
   wiggles will be integrated in the data of luminosity distances.
   A study shows that the model whose Hubble parameter
   is directly endowed with oscillating ansatz by parameterizations
   fits the data much better than those of LCDM, IntLCDM, XCDM, IntXCDM,
   VecDE, IntVecDE \cite{weihao2}. A physical model, in which the phantom field with natural potential
 ,i.e, the potential of a pseudo Nambu-Goldstone Boson (PNGB) plays the role of
 dark energy, is investigated in \cite{hongsheng}. The oscillating
 behavior of $H$ appears naturally in this model.

  For $\Lambda$CDM, in the joint analysis with a marginalization of $H_0$, $\chi^2$
  reads
  \be
 \chi^2(\Omega_{m0},\Omega_{k0})=\sum\limits_{i=1}^{6}\left[\frac{f_{obs}(z_i)-
 f_{th}(z_i;\Omega_{m0},\Omega_{k0})}{\sigma_{f_{obs}}}\right]^2+\sum\limits_{i=1}^{9}\left[\frac{H_{obs}(z_i)-
 H_{th}(z_i;\Omega_{m0},\Omega_{k0})}{\sigma_{f_{obs}}}\right]^2+\left(\frac{H_0-72}{0.08}\right)^2,
 \en
 where $f_{obs}$ denotes the observation value of the growth index,
 and $f_{th}$ represents its theoretical value.
 We read $f_{obs}(z_i)$, $\sigma_{f_{obs}}$ from Table 1 and
 calculate $f_{th}(z_i;\Omega_{m0},\Omega_{k0})$ using (\ref{diff}).
  To get the theoretical value of $f$ using (\ref{diff}),
 we need its initial value. Our considerations are as
 follows. In any dark energy model the universe should behave as the same one in some high
 redshift region such as $z=1000$, that is, it behaves as standard cold dark matter (SCDM) model,
 which has been sufficiently tested by observations. In SCDM model
 we obtain $f=1$ by using (\ref{f}). So we just take $f=1$ as the initial value at
 high enough
 redshift region. And the theoretical Hubble parameter reads,
 \be
 H_{th}^2=H_0^2\left[\Omega_{m0}(1+z)^3+\Omega_{k0}(1+z)^2+1-\Omega_{m0}-\Omega_{k0}\right].
 \en
 We take the value of present Hubble parameter $H_0$  from the HST key project $H_0= 0.72 \pm 0.08$kms$^{-1}$Mpc$^{-1}$ \cite{h0}.
  The result is shown in fig. 1.

  In DGP model, traditionally, we often use $\Omega_{r_c}$ rather than $\Omega_{k0}$ in
 fittings. There is no essential difference since they are
 constrained by (\ref{rcmk}). In the joint analysis with a marginalization of $H_0$, $\chi^2$
 becomes
  \be
 \chi^2(\Omega_{m0},\Omega_{r_c})=\sum\limits_{i=1}^{6}\left[\frac{f_{obs}(z_i)-
 f_{th}(z_i;\Omega_{m0},\Omega_{r_c})}{\sigma_{f_{obs}}}\right]^2+\sum\limits_{i=1}^{9}\left[\frac{H_{obs}(z_i)-
 H_{th}(z_i;\Omega_{m0},\Omega_{r_c})}{\sigma_{f_{obs}}}\right]^2+\left(\frac{H_0-72}{0.08}\right)^2,
 \en
 where
 \be
 H_{th}^2=H_0^2\left[\left((\Omega_{m0}(1+z)^3 + \Omega_{r_c})^{1/2} + \Omega_{r_c}^{1/2}\right)^2 +
 \Omega_{k0}(1+z)^2\right].
  \en
  With the same reason as the case of $\Lambda$CDM we take $f=1$
 as  the initial value at high enough  redshift. The result is
 illuminated in fig.2.

  \begin{figure}
 \centering
 \includegraphics[totalheight=2.8in, angle=0]{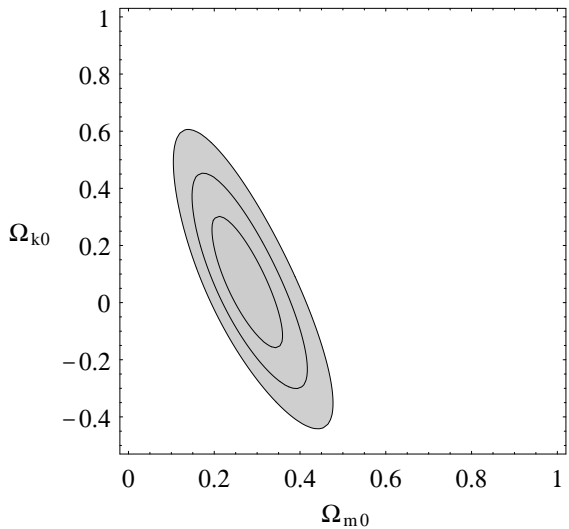}
 \caption{68\% , 95\% and 99\% confidence contour plot of $\Omega_{m0},~\Omega_{k0}$ in
 $\Lambda$CDM by the growth data in Table I and $H(z)$ data in Table
 II.
  For 1 $\sigma$ level, $\Omega_{m0}=0.275^{+0.0544}_{-0.0549}$,
 $\Omega_{k0}=0.065^{+0.159}_{-0.149}$. }
 \label{jointlcdm}
 \end{figure}

  \begin{figure}
 \centering
 \includegraphics[totalheight=2.8in, angle=0]{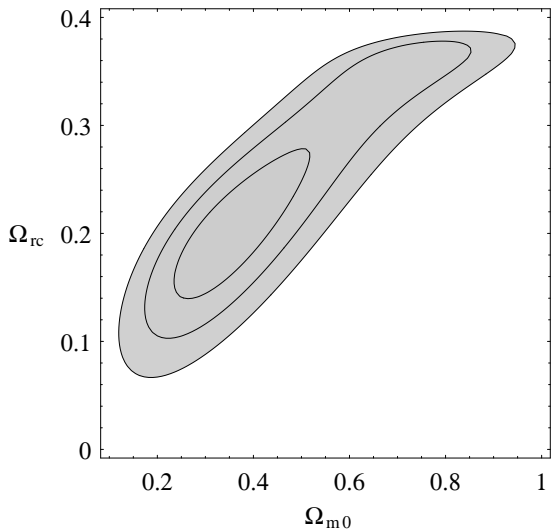}
 \caption{68\% , 95\% and 99\% confidence contour plot of $\Omega_{m0},~\Omega_{k0}$ in
 DGP by the growth data in Table I and $H(z)$ data in Table
 II.
  For 1 $\sigma$ level, $\Omega_{m0}=0.350^{+0.132}_{-0.0974}$,
 $\Omega_{r_c}=0.200^{+0.0631
 }_{-0.0483}$. }
 \label{jointdgp}
 \end{figure}

   Observing Table 1 carefully, one may find that the datum at $z=3.0$ is odd in some
   degree. From (\ref{f}) we see that in $\Lambda$CDM $\Omega_{k}$ or $\Omega_{de}$
   should be smaller than $0$ if we require $f>1$. Hence our present
   universe will be curvature dominated or becomes an anti-de
   Sitter (AdS) space, since dust matter redshifts much faster than
   curvature or vacuum energy. Here we give a simple exmple of this
   problem. In the spatially flat $\Lambda$CDM model,  $f=1.46$
   yields,
 \be
 \Omega_m^{6/11}=1.46.
 \en
 Then we derive $\Omega_m=2.00, \Omega_{de}=-1.00$. The universe will
 brake and then start to contract at $z=2.17$, which completely contradicts to the
  observations of expansion. So we present fig. 3, which displays the constraint on
  $\Omega_{m0},~\Omega_{k0}$ in $\Lambda$CDM by
  $H(z)$ data and growth data, which only include 5 points. Similarly, we plot fig. 4, which illuminates the constraint on
  $\Omega_{m0},~\Omega_{r_c}$ in DGP by
  $H(z)$ data and growth data, which only includes 5 points. The datum at $z=3.0$ is
  excluded.

    \begin{figure}
  \centering
  \includegraphics[totalheight=2.8in, angle=0]{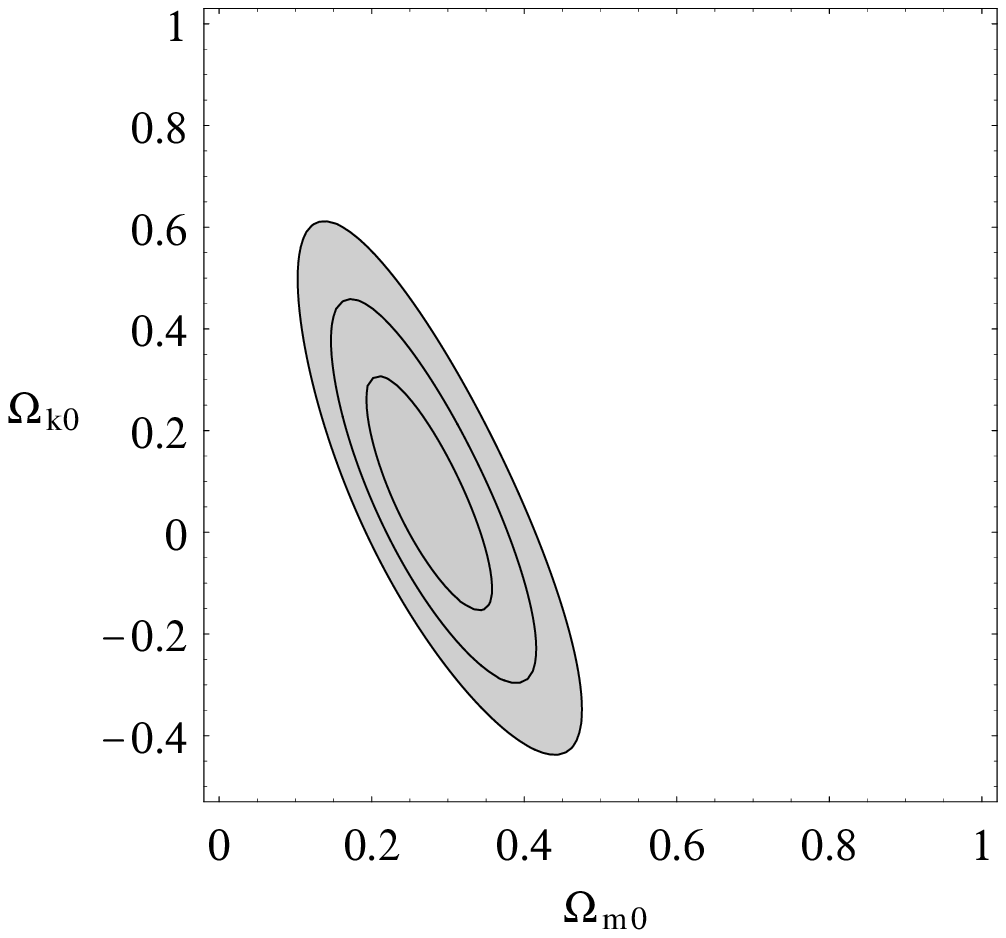}
   \caption{68\% , 95\% and 99\% confidence contour plot of $\Omega_{m0},~\Omega_{k0}$ in
  $\Lambda$CDM by the growth data in Table I and $H(z)$ data in Table
  II.
  For 1 $\sigma$ level, $\Omega_{m0}=0.270^{+0.0598}_{-0.0531}$,
  $\Omega_{k0}=0.080^{+0.146}_{-0.159}$. The point $z=3.0$ in the sample of the growth data is excluded. }
  \label{jointlcdm5}
  \end{figure}

  \begin{figure}
 \centering
 \includegraphics[totalheight=2.8in, angle=0]{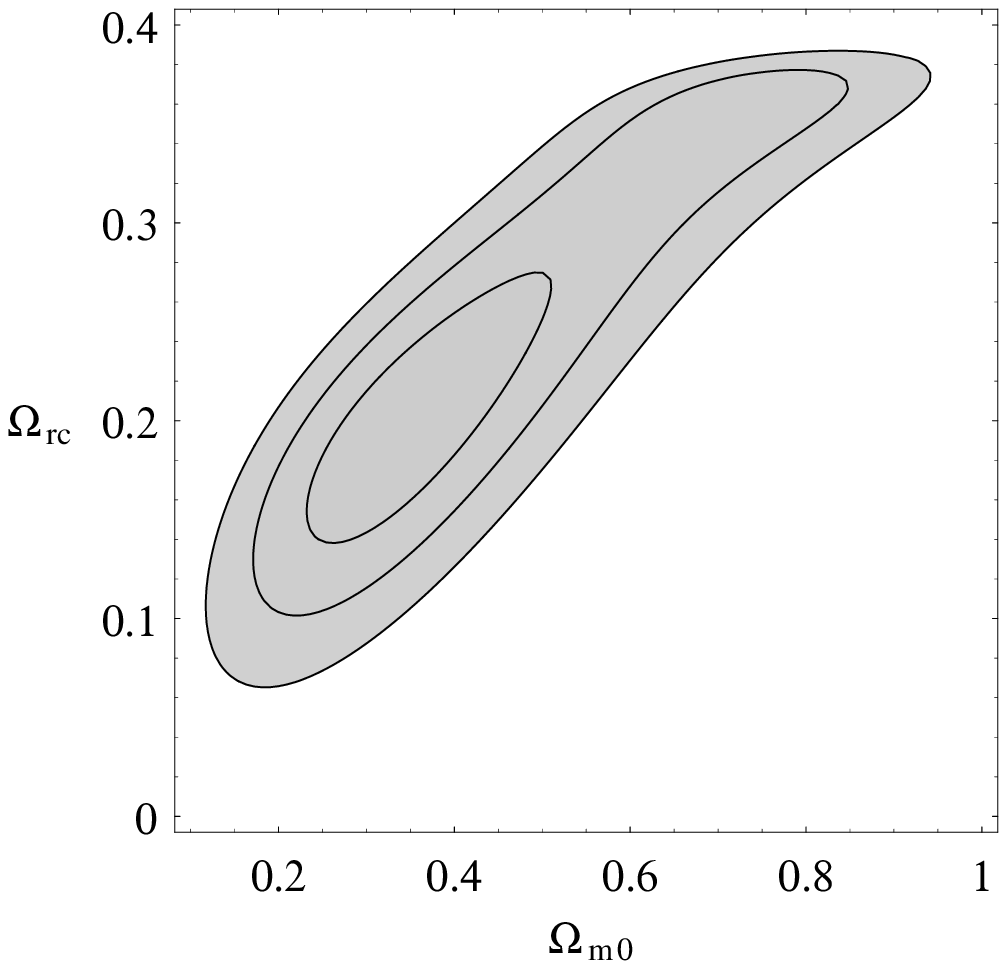}
 \caption{68\% , 95\% and 99\% confidence contour plot of $\Omega_{m0},~\Omega_{k0}$ in
 DGP by the growth data in Table I and $H(z)$ data in Table
 II.
  For 1 $\sigma$ level, $\Omega_{m0}=0.345^{+0.1296}_{-0.0940}$,
 $\Omega_{r_c}=0.198^{+0.0613}_{-0.0471}$. The point $z=3.0$ in the sample of the growth data is excluded.}
 \label{jointlcdm6}
 \end{figure}

    Comparing fig. 1 with fig 3., we find the profiles of the two figures are almost the same,
    but the minimal $\chi^2$, $\chi^2_{\rm min}$ decreases from $
    12.26$ to $ 9.479$. Similarly, comparing fig. 2 with fig. 4,
     we find the profiles of the two figures are almost the same,
    but  $\chi^2_{\rm min}$ decreases from $
    12.11$  to $ 9.375$. Without the point at $z=3.0$ the constraints on
    $\Omega_{k0}$ of $\Lambda$CDM, $\Omega_{m0}$ and $\Omega_{r_c}$ of DGP
    become more tighten instead. This is also a signal that the datum
    $z=3.0$ may not be well consistent with other data.

     If we only consider $\chi^2_{\rm min}$, we may
    conclude that DGP is more favored. However, $\chi^2_{\rm min}$
    is only one point and the difference is tiny between the two
    models. We need more comparisons with the independent results, especially the permitted parameter internals,
     fitted by the expansion data, which were
    thoroughly studied.
       The latest results are shown as
 follows. For $\Lambda$CDM model, $\Omega_{m0}=0.279\pm 0.008$,
 $\Omega_{k0}=-0.0045\pm 0.0065$, which are derived from the joint
 analysis of
 the CMB (five-year WMAP data), the distance measurements from the
 Type Ia SN, and the Baryon Acoustic Oscillations (BAO) in the distribution of galaxies \cite{WMAP}.
 For DGP model, $\Omega_{m0}=0.28^{+0.03}_{-0.02}$,
 $\Omega_{r_c}=0.13\pm 0.01$ (SN(new gold)+CMB+SDSS+gas), and  $\Omega_{m0}=0.21\pm 0.01$,
 $\Omega_{r_c}=0.16\pm 0.01$(SN(SNLS)+CMB+SDSS+gas) \cite{dgp2007}.

 For $\Lambda$CDM, the result of joint fitting by growth data and $H(z)$ data $\Omega_{m0}=0.275^{+0.0544}_{-0.0549}$,
 $\Omega_{k0}=0.065^{+0.159}_{-0.149}$, almost coincides with the
 result by expansion data $\Omega_{m0}=0.279\pm 0.008$,
 $\Omega_{k0}=-0.0045\pm 0.0065$. These types of data are well
 consistent in frame of $\Lambda$CDM model.

 For DGP, the result of joint fitting by growth data and $H(z)$ data impose $\Omega_{m0}=0.350^{+0.132}_{-0.0974}$,
 $\Omega_{r_c}=0.200^{+0.0631
 }_{-0.0483}$, which is not well consistent with the
 result by expansion data $\Omega_{m0}=0.28^{+0.03}_{-0.02}$,
 $\Omega_{r_c}=0.13\pm 0.01$ (SN(new gold)+CMB+SDSS+gas), and  $\Omega_{m0}=0.21\pm 0.01$,
 $\Omega_{r_c}=0.16\pm 0.01$(SN(SNLS)+CMB+SDSS+gas). Concretely speaking,
 $\Omega_{r_c}=0.200^{+0.0631
 }_{-0.0483}$(growth+$H(z)$) inhabits beyond $2\sigma$ level of
 expansion data $\Omega_{r_c}=0.13\pm 0.01$ (SN(new
 gold)+CMB+SDSS+gas). For the data set SN(SNLS)+CMB+SDSS+gas, the result by growth and $H(z)$
  $\Omega_{m0}=0.350^{+0.132}_{-0.0974}$ dwells beyond $3\sigma$
  level of $\Omega_{m0}=0.21\pm 0.01$. Therefore DGP model can not
  fit the observations of expansion and growth very well at the same time.

  Through the above discussions, we see that the dark energy model
  is more favored than the DGP model by the present data,
  and the growth data can be an  effective probe to study the nature of the dark energy.

 We plot the best fit
 curves of growth $f$  by growth+H(z) data and expansion
 data in $\Lambda$CDM, respectively in fig. 5. Fig. 6 illuminates
 the best fit curves by growth+H(z) data and expansion data in DGP. It is
 clear that the gap between the best fit curves of growth data and
 expansion data is much bigger in DGP model than the gap in
 $\Lambda$CDM model.

     \begin{figure}
 \centering
 \includegraphics[totalheight=2.8in, angle=0]{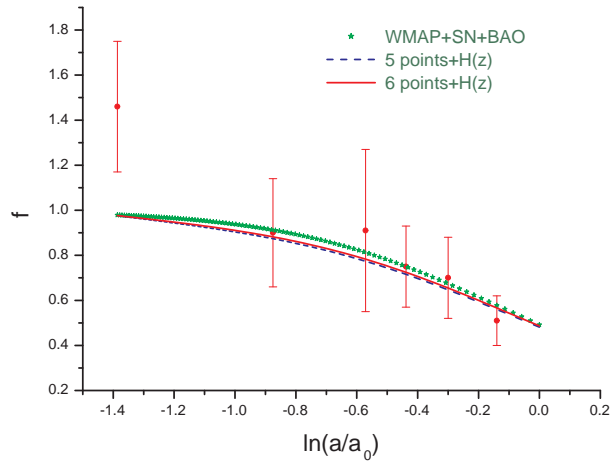}
 \caption{The perturbation growth $f$ with error bars in Table 1 and the best fit curves
 in $\Lambda$CDM model. The best fit result by growth data and $H(z)$ data inhabits on the red solid curve,
  the best fit result by growth data except the point $z=3.0$ and $H(z)$ data resides on the blue dashed one, and
  and the best fit result by joint analysis of WMAP, SN and BAO dwells on the green triangle ones.}
 \label{bestfitlcdm}
 \end{figure}

 \begin{figure}
 \centering
 \includegraphics[totalheight=2.8in, angle=0]{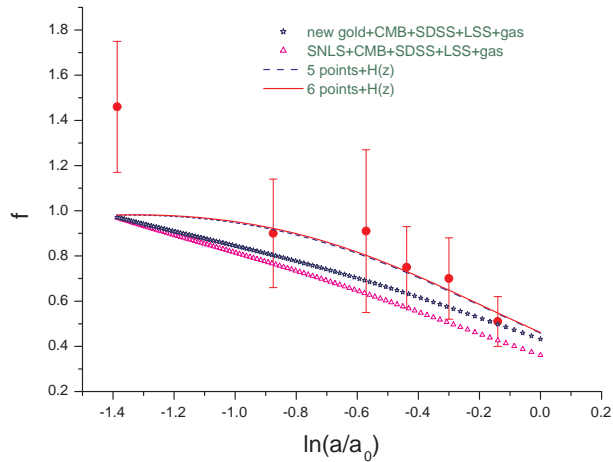}
 \caption{The perturbation growth $f$ with error bars in Table 1 and the best fit curves
 in DGP model. The best fit result by growth+H(z) data inhabits on the red solid curve,
  the best fit result by growth data except the point $z=3.0$ and $H(z)$ data resides on the blue dashed one,
  the best fit result by joint analysis of SN(new gold), CMB, SDSS, and gas dwells on the navy star curve,
  and the best fit result by joint analysis of SN(SNLS), CMB, SDSS, and gas is denoted by the pink triangle ones.}
 \label{bestfitdgp}
 \end{figure}

\section{Conclusions}
  Perturbation growth is a newly developed method to differentiate between
  dark energy and modified gravity. In the previous works people concentrate on the
  approximate analytical value of
  the perturbation growth index of a model, and then compare with
  the observations. But, the index is not a constant in the history
  of the universe. We fit dark energy and modified gravity models by
  using the exact evolution equation of perturbation growth.

  The sample of presently available growth data is quite small and
  the error bars are rather big. Furthermore, we always assumed
  $\Lambda$CDM model for deriving the growth data. Thus it seems
  proper to fit a model by jointing the growth data and the expansion
  data.  The direct $H(z)$ data are new type of data, which can be used to
  explore the fine structures of the Hubble expansion history.
  We put forward a joint fitting by the growth data and $H(z)$ data.
  The results are summarized as follows: For $\Lambda$CDM, $\Omega_{m0}=0.275^{+0.0544}_{-0.0549}$,
  $\Omega_{k0}=0.065^{+0.159}_{-0.149}$; For DGP, $\Omega_{m0}=0.350^{+0.132}_{-0.0974}$,
 $\Omega_{r_c}=0.200^{+0.0631 }_{-0.0483}$.

 The minimal $\chi^2$ are 12.26 and 12.11 for $\Lambda$CDM and DGP, respectively.
 The permitted parameters of $\Lambda$CDM by
growth+H(z) data show an excellent consistency with the previous
results inferred from expansion data. However, for DGP model the
discrepancies of the results of growth+H(z) data and expansion data
are at least 2$\sigma$ level. Hence 9 in the sense of consistency,
 $\Lambda$CDM is more favored than DGP.

 \section*{Acknowledgments}
 We thank the anonymous referee for
 several valuable suggestions. We thank Hao Wei for helpful
 discussions.
 H.Noh was supported by grant No. C00022 from the Korea Research
 Foundation. Z.H. Zhu was supported by
  the National Natural Science Foundation of China
    , under Grant No. 10533010, by Program for New Century Excellent
    Talents in University (NCET) and SRF for ROCS, SEM of China.

\end{document}